\documentclass[runningheads]{llncs}

\usepackage{graphicx}
\usepackage{booktabs}
\usepackage{subcaption}
\usepackage[accsupp]{axessibility}  

\usepackage{hyperref}

\usepackage{orcidlink}

\usepackage{float}
\usepackage{placeins}
\usepackage{amsmath, amssymb}
\begin{document}

\title{Weakly Conditioned Diffusion Model Augmentation for Prostate Segmentation} 

\titlerunning{Abbreviated paper title}

\author{Chibike Ugbam\inst{1}\orcidlink{0000-1111-2222-3333} \and
Muhammad Bilal\inst{2}\orcidlink{0000-0001-8446-7123}
\and
Shazad Ashraf\inst{3}
\and
Jia Wu\inst{4}\orcidlink{0000-0001-8392-8338}
\and
Shafaq Murad
\inst{5}
\and
Rizwan Qureshi
\inst{6, 7}
\and
Hazrat Ali\inst{1}\orcidlink{0000-0003-3058-5794}}

\authorrunning{C.~Ugbam et al.}

\institute{University of Stirling, United Kingdom.
\and
Birmingham City University, United Kingdom\\
\and
Department of Colorectal Surgery, Queen Elizabeth Hospital,\\ University Hospitals Birmingham NHS Foundation Trust, Edgbaston Birmingham, United Kingdom.
\and
MD Anderson Cancer Center, The University of Texas, USA.\\
\and
Manchester University NHS Foundation Trust, Manchester Royal Infirmary,\\ Oxford Road, Manchester, M13 9WL, United Kingdom.\\
\and
Massachusetts General Hospital, MGH, Harvard University, USA.
\and
Laboratory for Information and Decision Systems, Massachusetts Institute of Technology, Massachusetts Avenue Cambridge, MA, USA.\\
\email{c.o.ugbam@stir.ac.uk, muhammad.bilal@bcu.ac.uk, 
shazad.ashraf@uhb.nhs.uk,
JWu11@mdanderson.org, 
shafaq.murad@mft.nhs.uk, rqureshi3@mgh.harvard.edu, hazrat.ali@live.com}\\
}

\maketitle

\begin{abstract}
  Deep learning models for prostate MRI segmentation often struggle to generalize when trained on limited heterogeneous data, leading to boundary errors and over-segmentation. We present an augmentation framework using mask-conditioned synthetic and diverse T2-weighted MRI volumes to improve boundary precision and reduce over-segmentation. We employ a 2.5D weakly mask-conditioned denoising diffusion probabilistic model (DDPM) via channel concatenation. The diffusion model conditions on stacks of adjacent binary masks from training folds. The model generates T2 volumes consistent with the conditioning anatomy while diversifying tissue appearance. We evaluate the proposed approach on two cohorts of data: (i) the LUND probe cohort (n=432) for strategy selection and training nnU-Net from scratch with a 23\% synthetic data mix, and (ii) the Medical Segmentation Decathlon (MSD) prostate cohort (n=32) evaluated with a 20\% synthetic data mix. For the LUND cohort, we found that training an nnU-Net model from scratch with a 23\% synthetic data mix reduced the mean false-positive voxels per case from 11,102 (baseline) to 10,757. On the MSD cohort, the augmented model with a 20\% synthetic mix achieved a mean cross-validated Dice of 0.9094, improving over the baseline (0.8935) by 1.59 percentage points and reducing mean false-positive voxels by 23.1\% (4,201 to 3,229).
Experimental results on both these cohorts suggest that mask-conditioned DDPM augmentation can reduce over-segmentation errors and improve boundary adherence in data-constrained settings.
\end{abstract}

\section{Introduction}
\label{sec:intro}
Prostate cancer is one of the most prevalent cancers in men worldwide, with incidence projected to increase with age \cite{ sung2021}. T2-weighted MRI has become an integral part of the diagnostic and treatment planning pathway \cite{turkbey2019}. A key step in prostate cancer diagnosis and treatment planning is \textit{segmentation}. Deep learning models have shown considerable promise for automated segmentation of MRI data, but their clinical deployment remains constrained by data scarcity. High-quality annotated datasets are expensive to produce, and models trained on limited data frequently generate boundary errors and over-segmentation when tested on unseen data.

Synthetic data augmentation using generative models offers a potential solution to this problem. Denoising diffusion probabilistic models (DDPMs) have emerged as a compelling framework, offering stable training and diverse high-fidelity image synthesis \cite{ho2020}. Conditioning generation on segmentation masks further ensures that synthesized images respect anatomical structure while varying in appearance. This approach is well-suited to producing augmented training data that broadens the distribution and does not simply replicate individual cases. 

This work investigates whether a 2.5D weakly mask-conditioned DDPM, implemented via channel concatenation of binary prostate masks, can generate T2-weighted MRI volumes that improve boundary precision and reduce over-segmentation in a prostate segmentation pipeline. The model is trained on the LUND-PROBE dataset \cite{rogowski2025}, a recently released open benchmark for MRI-based prostate radiotherapy. We train an nnU-Net backbone \cite{isensee2021} on a controlled mix of real and synthetic data. We evaluate the model on two cohorts of data: the LUND-PROBE (n=432) for strategy selection and the Medical Segmentation Decathlon (MSD) prostate cohort (n=32) \cite{antonelli2022} for cross-validated evaluation of cross-domain augmentation transfer. The hypothesis is that a moderate synthetic data proportion, approximately 20-23\%, can regularise the segmentation model and improve generalization to unseen acquisition settings. Robust prostate segmentation has direct clinical relevance as more generalizable models reduce the annotation burden for multi-site deployment and support safer, more consistent radiotherapy planning.

\section{Related Work}

U-Net and its variants have often dominated volumetric medical image segmentation \cite{ronneberger2015}. The nnU-Net framework formalized this robust pipeline configuration, including preprocessing, normalization, post-processing, and inference, and demonstrated performance improvements beyond architectural novelty alone \cite{isensee2021}, making it a widely adopted baseline for volumetric medical image segmentation and the backbone of this work.
However, segmentation models trained using data from one institution degrade when evaluated at another in part due to differences in scanner, field strength, and patient demographics. The PROMISE12 challenge was an early benchmark that highlighted this problem for prostate MRI \cite{litjens2014}. More recently, Alqarni et al. used nnU-Net and found that scanner and protocol differences produce substantial performance gaps, with heterogeneous training data offering only partial relief \cite{alqarni2025}. These findings motivate the need for augmentation to achieve generalization rather than increasing data volume.

GAN-based augmentation has previously been used to address data scarcity in MRI \cite{huang2021}, \cite{chen-gan-2022}, but GANs are prone to mode collapse and anatomical distortions that are unacceptable in clinical contexts. In comparison, diffusion models offer improved sample diversity and training stability, and their growing adoption across medical imaging synthesis tasks has been well documented \cite{kazerouni2023}. Channel-concatenation conditioning is an established mechanism for weakly guided diffusion generation. Med-DDPM demonstrated that channel-wise concatenation of segmentation masks allows a DDPM to synthesize anatomically coherent brain MRI and that combining real and synthetic data improves downstream segmentation beyond GAN-based baselines \cite{dorjsembe2024}. This soft-conditioning strategy is adopted here as a computationally efficient mechanism that provides sufficient anatomical guidance while preserving the appearance diversity necessary for effective augmentation. The model is informed by the mask but not rigidly constrained by it, allowing it to generate varied tissue appearances from the same anatomical layout.
Konz et al. also demonstrated that segmentation mask-guided diffusion achieves strong anatomical faithfulness across breast MRI and abdominal CT synthesis tasks \cite{konz2024}. Pertinent to the present work is that Giardina et al. proposed a 3D anatomically guided diffusion model conditioned on prostate zone and lesion masks for synthetic prostate MRI generation, showing the potential of mask-conditioned diffusion as an augmentation strategy for cancer segmentation using the PI-CAI dataset \cite{giardina2025}. 

The present work differs in three important aspects: (1) we operate in 2.5D rather than 3D to allow development under computational constraints, (2) we use weaker channel concatenation conditioning to prioritize distributional diversity over synthesis fidelity, and (3) our DDPM is trained on a radiotherapy-specific dataset and its generative prior is transferred to a second domain via mask-conditioned synthesis, rather than evaluating synthesis quality within a single domain. Critically, prior works have not applied diffusion-based synthesis to the LUND-PROBE dataset, which is radiotherapy-specific and represents a clinically distinct and underexplored territory. 
Our work fills this gap by evaluating whether \textit{weakly mask-conditioned DDPM augmentation on LUND-PROBE can reduce over-segmentation and improve boundary adherence} in external validation, using synthetic augmentation as a generalization tool and not just for data scarcity.

\section{Methodology}
\subsection{Dataset} 
Two cohorts were used in this study. The primary dataset is LUND-PROBE \cite{rogowski2025}, a recently released large open benchmark for MRI-based radiotherapy comprising 432 T2-weighted MRI volumes with voxel-level prostate delineations acquired under realistic clinical conditions. The data set was partitioned into 302 training cases and 130 test cases.
The prostate subset of the MSD \cite{antonelli2022}, \cite{simpson2019} was used as a secondary cohort to evaluate cross-domain augmentation transfer, comprising 32 cases in total. Since the MSD challenge test labels are held by the organizers and are not publicly available, evaluation was performed using 5-fold cross-validation. Synthetic augmentation data for the MSD cohort was generated by transferring the LUND-PROBE generative model to MSD, but the MSD synthetic data pipeline remained isolated. No held-out MSD images were used during synthetic data generation.

While the MSD dataset is multiparametric, including T2-weighted (T2w) and apparent diffusion coefficient (ADC) modalities, this work utilized only the T2w modality. This decision was motivated by two reasons. First, T2w imaging is the clinical gold standard for delineating prostate zonal anatomy and offers higher in-plane resolution than ADC. Second, restricting generation to a single high-resolution modality allowed the 2.5D DDPM to learn a more stable generative prior, avoiding the confounding effect of the lower resolution and different noise characteristics inherent to ADC maps.

\begin{figure}[!t]
\centering
\includegraphics[width=0.9\textwidth]{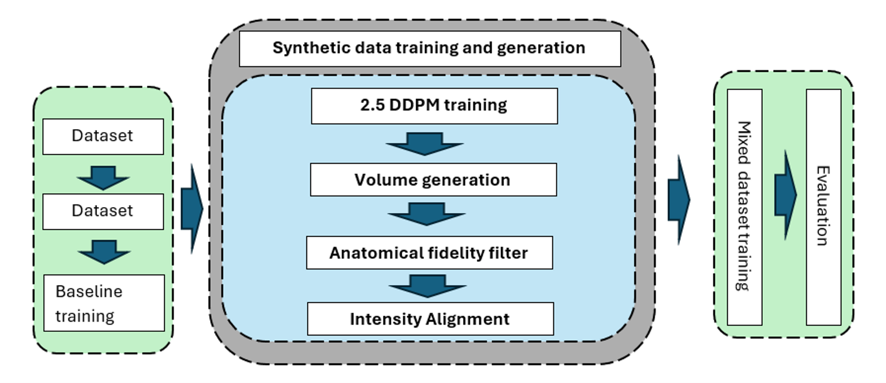}
\caption{Proposed methodology pipeline mapping the integration of synthetic data into nnU-Net training. Strategy selection is performed on the LUND-PROBE cohort prior to external validation on the MSD cohort.}
\label{fig:pipeline}
\end{figure}

\subsection{Experiment pipeline and training}
As shown in Figure~\ref{fig:pipeline}, we trained an nnU-Net \cite{isensee2021} model from scratch on the 302 LUND-PROBE training cases using the default self-configuring pipeline. We trained a separate nnU-Net baseline on the 32 MSD cases within the cross-validation framework to establish a domain-specific reference. In both cases, performance was evaluated using the Dice Similarity Coefficient (Dice) \cite{litjens2014} and mean false-positive (FP) voxels per case after standard nnU-Net post-processing, with false-positive voxels serving as a direct measure of over-segmentation.

\begin{figure}[!t]
\centering
\includegraphics[width=0.9\textwidth]{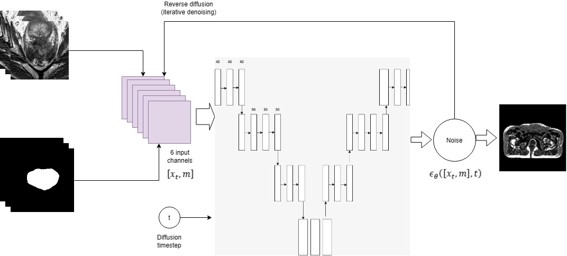}
\caption{Architecture of the proposed 2.5D mask-conditioned DDPM. Three adjacent T2-weighted MRI slices and their corresponding prostate masks form a six-channel input. A timestep embedding conditions the U-Net on the diffusion step, while the encoder-decoder with skip connections iteratively de-noises during reverse diffusion to generate synthetic MRI slice conditioned on the prostate mask.}
\label{fig:architecture}
\end{figure}

We trained the 2.5D DDPM \cite{ho2020} on axial slabs extracted from the LUND-PROBE training set (Figure \ref{fig:architecture}). The forward diffusion process progressively corrupts a clean image 
$\mathbf{x}_0$ over $T$ timesteps according to:

$$q(\mathbf{x}_t \mid \mathbf{x}_0) = 
\mathcal{N}\left(\mathbf{x}_t;\, \sqrt{\bar{\alpha}_t}\,\mathbf{x}_0,\, 
(1 - \bar{\alpha}_t)\mathbf{I}\right)$$

The model learns to reverse this process conditioned on the binary prostate mask $\mathbf{m}$, concatenated channel-wise with the noisy image at each denoising step:

$$p_\theta(\mathbf{x}_{t-1} \mid \mathbf{x}_t, \mathbf{m}) = 
\mathcal{N}\left(\mathbf{x}_{t-1};\, 
\mu_\theta\!\left([\mathbf{x}_t \,\|\, \mathbf{m}],\, t\right),\, 
\Sigma_\theta\!\left([\mathbf{x}_t \,\|\, \mathbf{m}],\, t\right)\right)$$

The network is trained by minimizing the simplified noise prediction objective:

$$\mathcal{L} = \mathbb{E}_{t,\, \mathbf{x}_0,\, \boldsymbol{\epsilon},\, \mathbf{m}}
\left[\left\|\boldsymbol{\epsilon} - 
\boldsymbol{\epsilon}_\theta\!\left([\mathbf{x}_t \,\|\, \mathbf{m}],\, 
t\right)\right\|^2\right]$$ 

\noindent where: 

$\mathcal{L}$ is the simplified variational lower bound objective minimized during training. 

$\mathbf{x}_0$ is the original clean T2-weighted image slab comprising a 3-slice axial stack, corrupted progressively during the forward process; $\mathbf{x}_t$ is the noisy image at diffusion timestep $t \in \{1, \ldots, T\}$, with $T=1000$, and $\mathbf{x}_{t-1}$ is the partially denoised image at the preceding timestep, which the reverse process learns to recover from $\mathbf{x}_t$.
    
$\mathbf{I}$ is the identity matrix that scales the noise variance in the forward process.

$\bar{\alpha}_t = \prod_{s=1}^{t}(1-\beta_s)$ is the cumulative noise schedule product, where $\beta_s$ is the variance of noise added at step $s$.

$\mathbf{m}$ is the binary prostate segmentation mask, concatenated channel-wise with $\mathbf{x}_t$ to form the 
conditioned input $[\mathbf{x}_t \,\|\, \mathbf{m}]$, constituting the weak conditioning mechanism - the mask 
informs but does not rigidly constrain the generative process. 

$\boldsymbol{\epsilon}$ is the Gaussian noise sampled during 
the forward process, whose prediction is the training target of 
the network $\boldsymbol{\epsilon}_\theta$, and $\boldsymbol{\epsilon}_\theta(\cdot)$ is the U-Net noise 
prediction network, where $\theta$ denotes its learnable parameters, 
trained to estimate $\boldsymbol{\epsilon}$ from the concatenated input $[\mathbf{x}_t \,\|\, \mathbf{m}]$ at timestep $t$. 

$\mu_\theta$ and $\Sigma_\theta$ are the predicted mean and variance of the reverse process, derived from the noise prediction network output.

$\mathbb{E}_{t,\,\mathbf{x}_0,\,\boldsymbol{\epsilon},\,\mathbf{m}}$ denotes the expectation taken over diffusion timesteps $t$, training images $\mathbf{x}_0$, sampled noise 
$\boldsymbol{\epsilon}$, and conditioning masks $\mathbf{m}$.

Each slab consisted of a 3-slice stack - a central slice and its two immediate neighbors - concatenated channel-wise with the corresponding binary prostate segmentation masks \cite{chen2024madm}. This channel concatenation constitutes the weak conditioning mechanism: the mask provides anatomical guidance without rigidly constraining generated appearance, allowing the model to diversify tissue contrast and texture while broadly respecting prostate boundaries. The model was implemented with a U-Net backbone with 48 base channels and trained with 1,000 diffusion timesteps, a maximum of 1,600 epochs with early stopping (patience=6, evaluated every 1,000 iterations), a batch size of 2 with gradient checkpointing, and a learning rate of $1 \times 10^{-4}$ using the AdamW optimizer without a dynamic learning rate scheduler. 

We generated synthetic T2-weighted prostate MRI volumes by conditioning the proposed DDPM on manually annotated prostate masks from the LUND-PROBE dataset. For external validation on the MSD cohort, the same DDPM, trained exclusively on LUND-PROBE data, was conditioned using MSD prostate masks. Consequently, synthetic MSD images inherited the anatomical geometry of the MSD masks while preserving the T2-weighted appearance characteristics learned from the LUND-PROBE training distribution. No real MSD image intensities were used during the synthesis process.

We assessed the quality of generated samples before downstream augmentation. Although distribution-level metrics such as the Fréchet Inception Distance (FID)~\cite{heusel2017} and its medical adaptations~\cite{woodland2024} are widely used to evaluate generative models, they are designed to measure global similarity between datasets rather than the fidelity of individual synthetic samples. Furthermore, conventional FID relies on feature representations extracted from networks trained on natural images, which are not optimized to characterize the intensity distributions and anatomical structures present in T2-weighted prostate MRI. For conditional medical image synthesis, where each synthetic image is generated from a known anatomical mask, per-sample assessment of both image realism and anatomical consistency is more appropriate than population-level distribution matching.

To address this, we introduce an anatomically informed composite filtering criterion for synthetic sample selection. Rather than evaluating generated images solely on perceptual similarity, the proposed criterion jointly considers image appearance and anatomical agreement. Specifically, the composite score is defined as:
\[
C(x,y)=\alpha\,\mathrm{SSIM}(x,y)+\beta\,\mathrm{JS}(P_x\parallel P_y),
\]

where $\mathrm{SSIM}(x,y)$ denotes the Structural Similarity Index Measure (SSIM) between the real image $x$ and the generated image $y$, and ${JS}(P_x\parallel P_y)$ is the Jensen–Shannon divergence computed between the intensity distributions of the real image $x$ and generated image $y$, restricted to the prostate region. The weighting coefficients were empirically fixed to $\alpha=0.85$ and $\beta=0.15$, prioritizing perceptual realism while introducing an explicit anatomical consistency constraint during sample selection. The Jenson-Shannon component acts as a geometric quality-control mechanism, reducing the likelihood that visually plausible but anatomically inconsistent synthetic samples are included in the augmented training set.

SSIM \cite{wang2004} is computed between each synthetic volume and its paired real source image, measuring structural fidelity at the voxel level. JS \cite{lin1991} is computed between the intensity distributions in the foreground of the synthetic and real images, restricted to the prostate region, penalizing the drift of intensity within clinically relevant tissue. Prior to computation, all slices were intensity-clipped at the 1st–99th percentile and min-max normalized, stabilizing the dynamic range of both metrics. Together, the formulation filters for samples that are structurally faithful and domain-matched in tissue intensity, while remaining diverse enough to broaden the training distribution. Only the top 30\% of generated samples were retained, yielding 90 synthetic volumes from LUND-PROBE and 8 from MSD.

Following filtering, histogram matching aligned retained synthetic intensities to real training cases in each cohort
Mixed training sets were then constructed at different synthetic ratios: 9\%, 18\%, 23\% synthetic mix, and 20\% synthetic mix for the MSD. In both settings, nnU-Net was trained from scratch under identical configurations to the respective baselines. LUND-PROBE performance was evaluated on the held-out test set (n=130); MSD performance was evaluated via 5-fold cross-validation. Primary metrics were mean DSC and mean false-positive voxels per case. Robustness was further assessed on 20\% of LUND-PROBE edge cases.

\section{Experiments and Results}

\subsection{LUND-PROBE Evaluation}
Five training configurations were evaluated using the nnU-Net self-configuring framework: (i) a baseline trained exclusively on real images, (ii) two SSIM-based synthetic augmentation strategies incorporating 9\% and 23\% synthetic data, and (iii) two configurations employing the proposed composite filtering criterion with 9\% and 18\% synthetic augmentation. All models were trained from scratch under identical training settings to ensure a fair comparison. Segmentation performance was evaluated on the independent LUND-PROBE test set ($n=130$) using Dice, the 95th percentile Hausdorff Distance (HD95), and false-positive and false-negative voxel counts.

\begin{table}[H]
\centering
\caption{Segmentation performance on the LUND-PROBE test set ($n=130$). FP and FN denote the mean number of false-positive and false-negative voxels per case.}
\label{tab:test_results}
\small
\setlength{\tabcolsep}{4pt}
\begin{tabular}{lcccc}
\hline
\textbf{Model} &
\textbf{Dice $\uparrow$} &
\textbf{HD95 $\downarrow$} &
\textbf{FP $\downarrow$} &
\textbf{FN $\downarrow$} \\
\hline
Baseline & 0.8984 & 3.5795 & 11102 & 10432 \\
SSIM-9\% & 0.8988 & 3.5685 & \textbf{10518} & 10873 \\
SSIM-23\% & 0.8989 & \textbf{3.5625} & 10757 & \textbf{10693} \\
Composite-9\% & \textbf{0.8990} & 3.5650 & 10612 & 10777 \\
Composite-18\% & 0.8988 & 3.5669 & 10532 & 10851 \\
\hline
\end{tabular}
\end{table}

Table~\ref{tab:test_results} summarises the performance of all training configurations. Across all models, mean Dice scores ranged from 0.8984 to 0.8990, while mean HD95 values ranged from 3.5625 mm to 3.5795 mm, demonstrating highly consistent segmentation performance irrespective of the augmentation strategy. All synthetic augmentation configurations reduced the mean false-positive voxel count relative to the baseline, indicating that the inclusion of carefully filtered synthetic images reduced over-segmentation while maintaining comparable false-negative rates. The Composite-9\% configuration achieved the highest mean Dice score (0.8990), whereas the SSIM-23\% configuration obtained the lowest mean HD95 (3.5625 mm). Although these numerical differences were small, they indicate that both SSIM-based and composite-filtered augmentation preserved segmentation performance while maintaining comparable error characteristics to the real-data baseline.

Although all synthetic augmentation configurations reduced mean false-positive voxels relative to the real-only baseline, each configuration also increased mean false-negative voxels. The results indicate a change in the false-positive/false-negative error profile rather than an unambiguous reduction in segmentation error. Importantly, the mean Dice and HD95 differences were small and none of the pairwise comparisons was statistically significant. Therefore, the false-positive reduction should not be interpreted in isolation as evidence of clinically improved contouring.

\begin{figure}[t]
\centering
\includegraphics[width=0.5\linewidth]{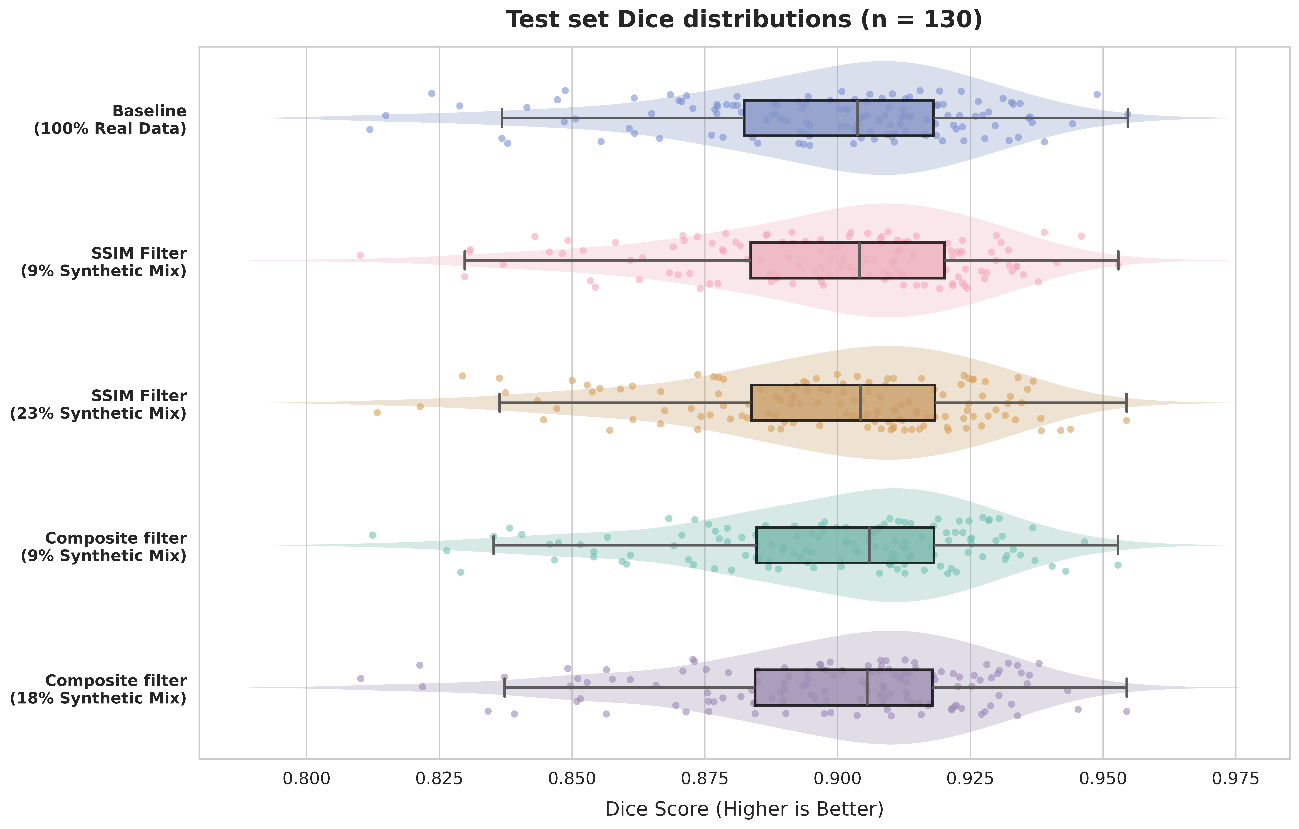}
\includegraphics[width=0.5\linewidth]{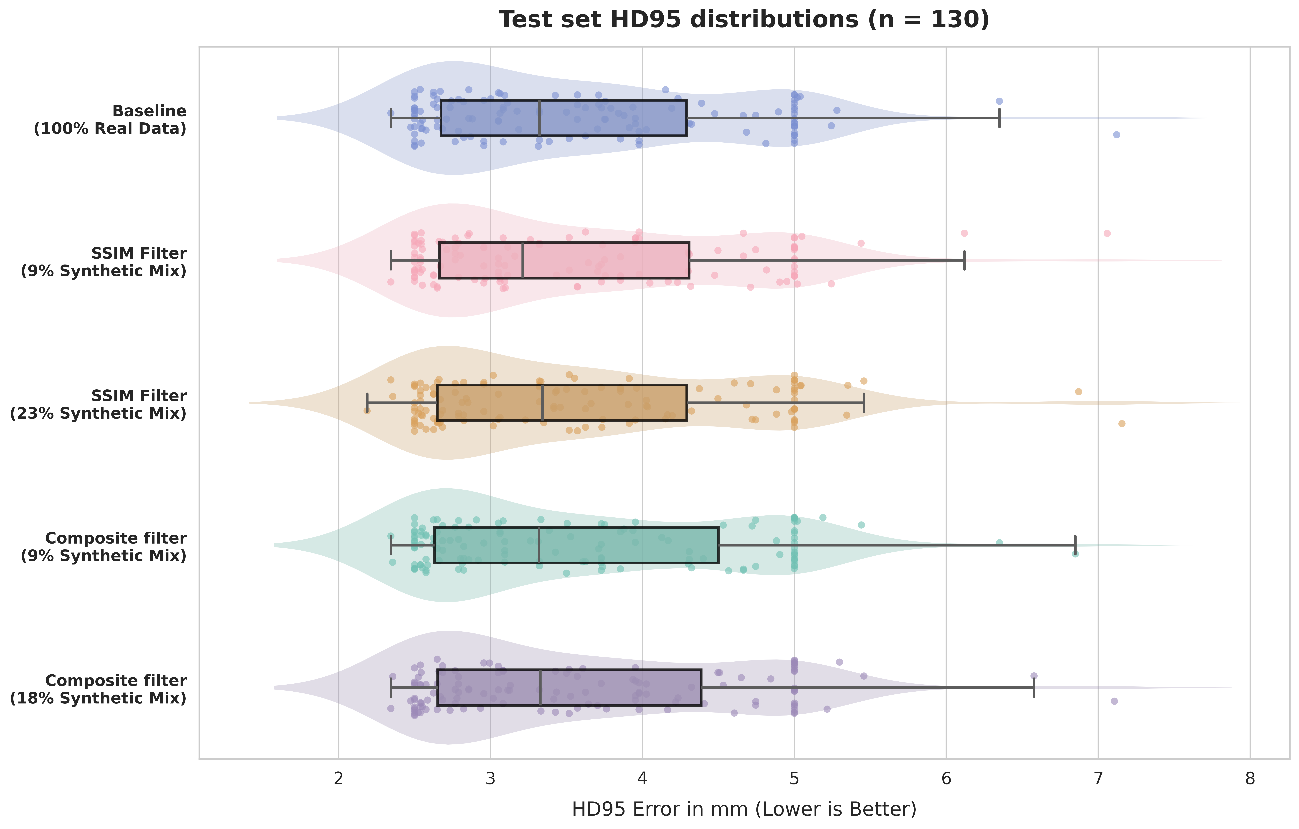}
\caption{Raincloud distributions of Dice and HD95 on the LUND-PROBE test set. Each raincloud combines the kernel density estimate, boxplot, and individual patient observations. The substantial overlap between all configurations indicates that synthetic augmentation preserved the overall segmentation performance and boundary accuracy of the baseline model.}
\label{fig:raincloud}
\end{figure}

Figure~\ref{fig:raincloud} presents the raincloud distributions of Dice and HD95 across all test cases. Considerable overlap between the distributions demonstrates that the synthetic augmentation strategies preserved the overall performance characteristics of the baseline model. Dice scores remained tightly centred around 0.90 for all configurations, while HD95 exhibited similar distributions across models, indicating that synthetic augmentation neither introduced additional variability nor degraded boundary accuracy.

\FloatBarrier

\subsection{Statistical analysis and Ablation study}

To determine whether the observed numerical differences represented meaningful performance changes, pairwise Wilcoxon signed-rank tests were performed for both Dice and HD95 (Table~\ref{tab:wilcoxon}). No statistically significant differences were observed between the baseline model and any synthetic augmentation strategy ($p>0.05$ for all comparisons). These results demonstrate that replacing a proportion of the training data with carefully selected synthetic images preserved segmentation performance.

\begin{table}[H]
\centering
\caption{Pairwise Wilcoxon signed-rank tests comparing segmentation performance between training configurations on the LUND-PROBE test set. Statistical significance was assessed at $p<0.05$.}
\label{tab:wilcoxon}
\small
\setlength{\tabcolsep}{4pt}
\begin{tabular}{lcc}
\hline
\textbf{Comparison} &
\textbf{Dice $p$-value} &
\textbf{HD95 $p$-value} \\
\hline
Baseline vs. SSIM-9\% & 0.2448 & 0.3176 \\
Baseline vs. SSIM-23\% & 0.1608 & 0.1528 \\
Baseline vs. Composite-9\% & 0.2612 & 0.0736 \\
Baseline vs. Composite-18\% & 0.3028 & 0.3627 \\
SSIM-9\% vs. Composite-9\% & 0.7423 & 0.2067 \\
Composite-9\% vs. Composite-18\% & 0.9787 & 0.4895 \\
\hline
\end{tabular}
\end{table}
\FloatBarrier

To evaluate the effect of the proposed anatomically informed filtering criterion independently of synthetic data proportion, an ablation study compared the SSIM - 9\% directly with the Composite - 9\%. No statistically significant differences were observed for either Dice ($p=0.7423$) or HD95 ($p=0.2067$), indicating that the introduction of an explicit anatomical consistency constraint during synthetic sample selection does not adversely affect downstream segmentation performance. Although the composite criterion did not significantly improve segmentation accuracy relative to SSIM filtering, it maintained equivalent performance while providing an additional mechanism for anatomical quality control during synthetic data curation.

A second ablation investigated the influence of synthetic data proportion by comparing the Composite - 9\% and Composite - 18\% configurations. Again, no statistically significant differences were observed for Dice ($p=0.9787$) or HD95 ($p=0.4895$), suggesting that increasing the proportion of composite-filtered synthetic data within the investigated range neither improved nor degraded segmentation performance. Collectively, these experiments demonstrate that the proposed filtering framework remains stable across moderate synthetic augmentation ratios.

\subsection{Worst-case robustness analysis}
Average performance metrics may obscure behaviour on clinically challenging cases. To investigate whether synthetic augmentation preferentially benefited difficult examples, an exploratory robustness analysis was performed using the lowest-performing 20\% of baseline Dice scores.

\begin{table}[H]
\centering
\caption{Worst-case robustness analysis performed on the lowest-performing 20\% of baseline Dice scores ($n=26$). Positive $\Delta$Dice indicates improvement relative to the control model, while negative $\Delta$HD95 indicates improved boundary accuracy. Statistical significance was assessed using paired Wilcoxon signed-rank tests.}
\label{tab:worst20}
\small
\setlength{\tabcolsep}{3.5pt}
\begin{tabular}{lcccc}
\hline
\textbf{Comparison} &
\textbf{$\Delta$Dice} &
\textbf{$p$} &
\textbf{$\Delta$HD95 (mm)} &
\textbf{$p$} \\
\hline
Baseline vs. SSIM-9\% & +0.0023 & 0.157 & $-$0.0535 & 0.117 \\
Baseline vs. SSIM-23\% & \textbf{+0.0029} & \textbf{0.022} & $-$0.0032 & 0.925 \\
Baseline vs. Composite-9\% & +0.0026 & 0.056 & $-$0.0466 & 0.071 \\
Baseline vs. Composite-18\% & +0.0011 & 0.367 & $-$0.0446 & 0.345 \\
SSIM-9\% vs. Composite-9\% & +0.0003 & 0.881 & +0.0069 & 0.689 \\
Composite-9\% vs. Composite-18\% & $-$0.0015 & 0.181 & +0.0020 & 0.730 \\
\hline
\end{tabular}
\end{table}
\FloatBarrier

For the lowest-performing 20\% of baseline cases, all synthetic augmentation configurations produced a positive numerical change in Dice relative to the baseline, with the largest improvement observed for SSIM-23\% (mean $\Delta$Dice = +0.0029, $p=0.022$). However, these cases were selected according to baseline performance, so the analysis is susceptible to regression to the mean and should be interpreted as exploratory rather than as an unbiased assessment of robustness. Therefore, the results suggest a possible benefit of synthetic augmentation on some challenging cases, but do not establish a robustness-specific improvement.

These findings suggest that the principal benefit of synthetic augmentation lies in improving robustness for difficult segmentation cases rather than increasing average-case performance. Because the difficult-case subsets were defined using baseline performance, these analyses should be interpreted as exploratory; however, the consistent trend across both subsets supports the hypothesis that synthetic augmentation primarily benefits anatomically challenging cases.

\subsection{Cross-validation on MSD Prostate Cohort}
The selected synthetic augmentation strategy was subsequently evaluated on the MSD prostate cohort using 5-fold cross-validation to assess its generalisability to an independent dataset. In each fold, nnU-Net was trained from scratch on the mixed dataset, comprising the real training fold cases supplemented with synthetic volumes, and evaluated on the held-out real validation cases. Table~\ref{tab:msd_results} reports the mean performance averaged across all five folds.

\begin{table}[h]
\centering
\caption{Segmentation performance on the MSD Prostate dataset using 5-fold cross-validation.}
\label{tab:msd_results}
\small
\begin{tabular}{lcc}
\hline
\textbf{Model} & \textbf{Mean Dice $\uparrow$} & \textbf{false-positive Voxels $\downarrow$} \\
\hline
Baseline (Real Only) & 0.8935 & 4,201.91 \\
Augmented (20\% Synthetic Mix) & 0.9094 & 3,229.90 \\
\hline
Net Improvement & +0.0159 (+1.59 pp) & $-$972.01 ($-$23.13\%) \\
\hline
\end{tabular}
\end{table}

The augmented model achieved a mean Dice score of 0.9094, an improvement of 1.59 percentage points over the real-only baseline, while reducing the mean false-positive voxel count by 972 voxels per case, corresponding to a 23.13\% reduction. These results are notable because the DDPM generative prior was trained exclusively on LUND-PROBE and was not trained on MSD image intensities. Instead, the learned LUND-PROBE prior was transferred to the MSD cohort by conditioning it on MSD prostate masks to generate synthetic T2-weighted images for augmentation. The results indicate that a generative prior learned from LUND-PROBE can be applied through mask-conditioned synthesis to augment segmentation training on the MSD cohort, with the augmented model achieving higher mean Dice and lower mean false-positive voxel counts in the five-fold cross-validation evaluation.

\section{Discussion}
The main finding of this study is that mask-conditioned synthetic augmentation preserved segmentation performance on the LUND-PROBE dataset while substantially improving generalization on an independent prostate MRI cohort. Although no statistically significant differences were observed in average Dice or HD95 across the LUND-PROBE test set, all synthetic augmentation strategies achieved segmentation performance comparable to the real-data baseline while consistently reducing false-positive predictions. Furthermore, the exploratory edge case analysis indicates that some challenging cases may benefit from synthetic augmentation. However, because the subset was defined using baseline Dice, this analysis is susceptible to regression to the mean and should not be interpreted as definitive evidence of a robustness-specific benefit.

The most clinically relevant result was the 23.13\% reduction in false-positive voxels on the MSD cohort. Because the DDPM was trained exclusively on LUND-PROBE data and never observed real MSD images during synthesis, this improvement demonstrates the successful transfer of a learned generative prior across acquisition domains. Rather than reproducing characteristics specific to the training dataset, the synthetic images introduced anatomically plausible variation that improved segmentation performance on previously unseen data.

The proposed composite filtering criterion was designed to address a limitation of generic image quality metrics for medical image synthesis. Metrics such as the Fréchet Inception Distance (FID), including medical adaptations, primarily assess distribution-level similarity and are not well suited to evaluating whether an individual synthetic image faithfully represents its conditioning anatomy. In contrast, the proposed composite score combines structural similarity with anatomical overlap, allowing synthetic images to be assessed on a per-sample basis using the image-mask pair available during generation. Although the composite filtering strategy did not produce statistically significant improvements over SSIM filtering alone, it achieved equivalent downstream segmentation performance while providing an additional anatomical consistency constraint during synthetic data selection.

Several limitations should be acknowledged. First, the MSD evaluation employed 5-fold cross-validation across 32 cases rather than a completely independent held-out test set. Second, the proposed diffusion model generates images in a 2.5D framework and therefore does not explicitly model long-range three-dimensional anatomical context. Finally, the weighting of the composite filtering criterion (85\% SSIM and 15\% Jensen-Shannon Index) was selected empirically rather than through formal optimization. Future work will investigate learned weighting strategies, volumetric diffusion models, and evaluation across larger multi-center datasets.

\section{Conclusion}

This study demonstrates that weakly mask-conditioned diffusion models can generate anatomically faithful synthetic T2-weighted prostate MRI suitable for data augmentation in medical image segmentation. Across the LUND-PROBE benchmark, synthetic augmentation preserved segmentation accuracy while maintaining performance statistically equivalent to the real-data baseline. The proposed composite filtering criterion further incorporated anatomical consistency into synthetic sample selection without compromising downstream segmentation performance.

Most importantly, synthetic augmentation substantially improved cross-dataset generalization. When applied to the independent MSD prostate cohort, the augmented model increased the mean Dice score by 1.59 percentage points while reducing false-positive voxels by 23.13\%, despite the diffusion model never being trained on MSD images. These findings demonstrate that anatomically guided synthetic augmentation can improve robustness and generalization for prostate MRI segmentation under limited data availability, supporting its potential as a practical strategy for clinical deep learning applications.


%

%

\appendix
\section{Qualitative comparison}
\begin{figure}[htbp]
     \centering
     \begin{subfigure}[b]{0.6\textwidth}
         \centering
         \includegraphics[width=\textwidth]{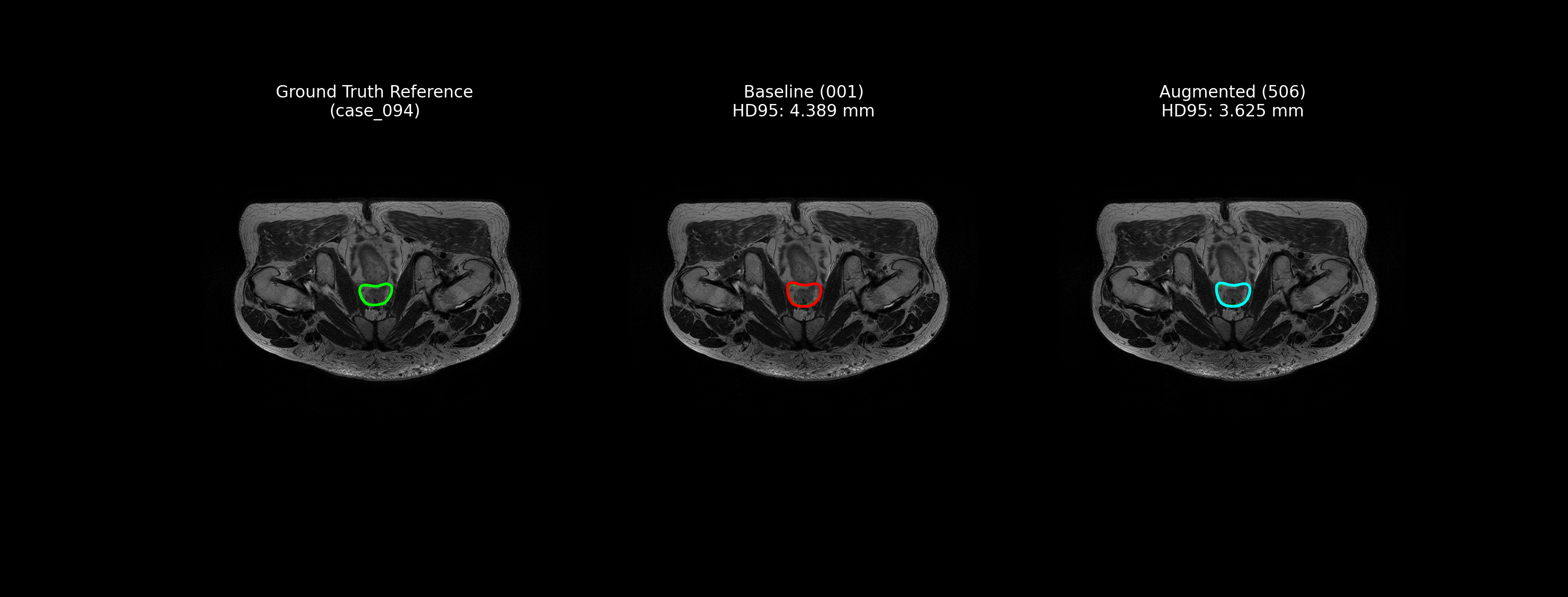}
         \caption{Case 094}
         \label{fig:hd95_case094}
     \end{subfigure}
      
     \begin{subfigure}[b]{0.6\textwidth}
         \centering
         \includegraphics[width=\textwidth]{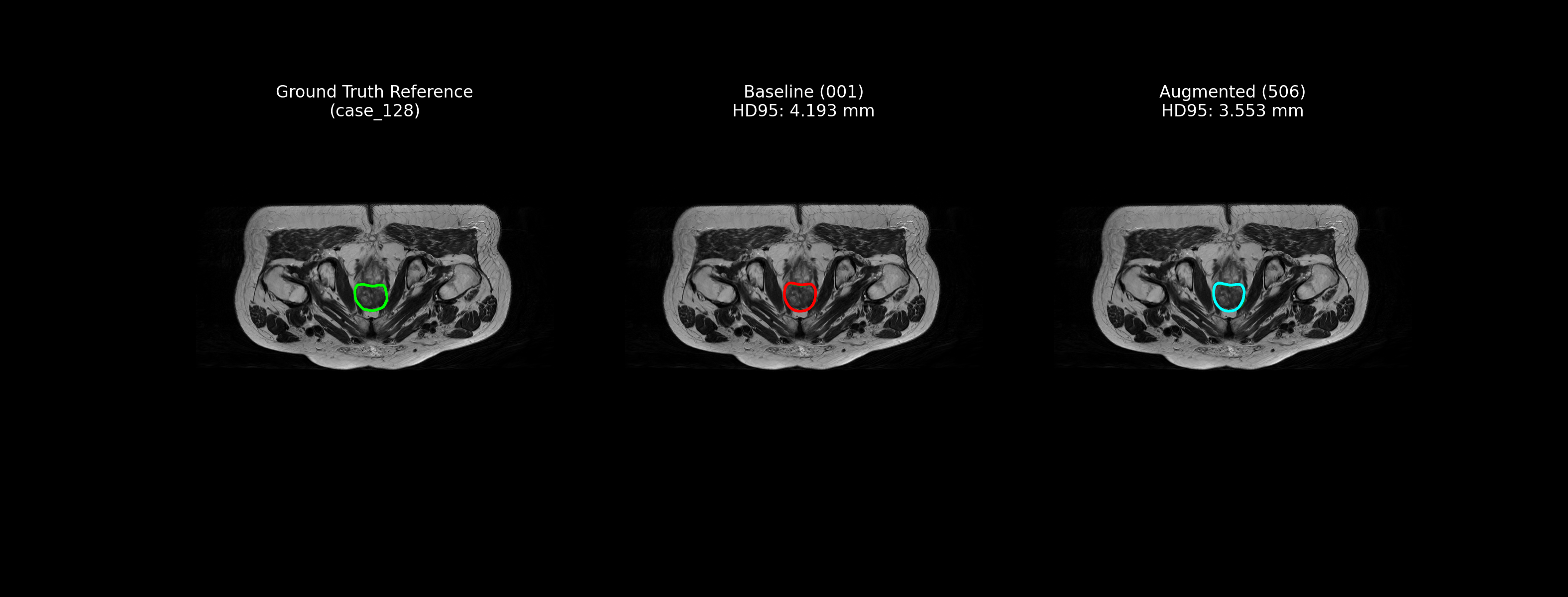}
         \caption{Case 128}
         \label{fig:hd95_case128}
     \end{subfigure}
      
     \begin{subfigure}[b]{0.6\textwidth}
         \centering
         \includegraphics[width=\textwidth]{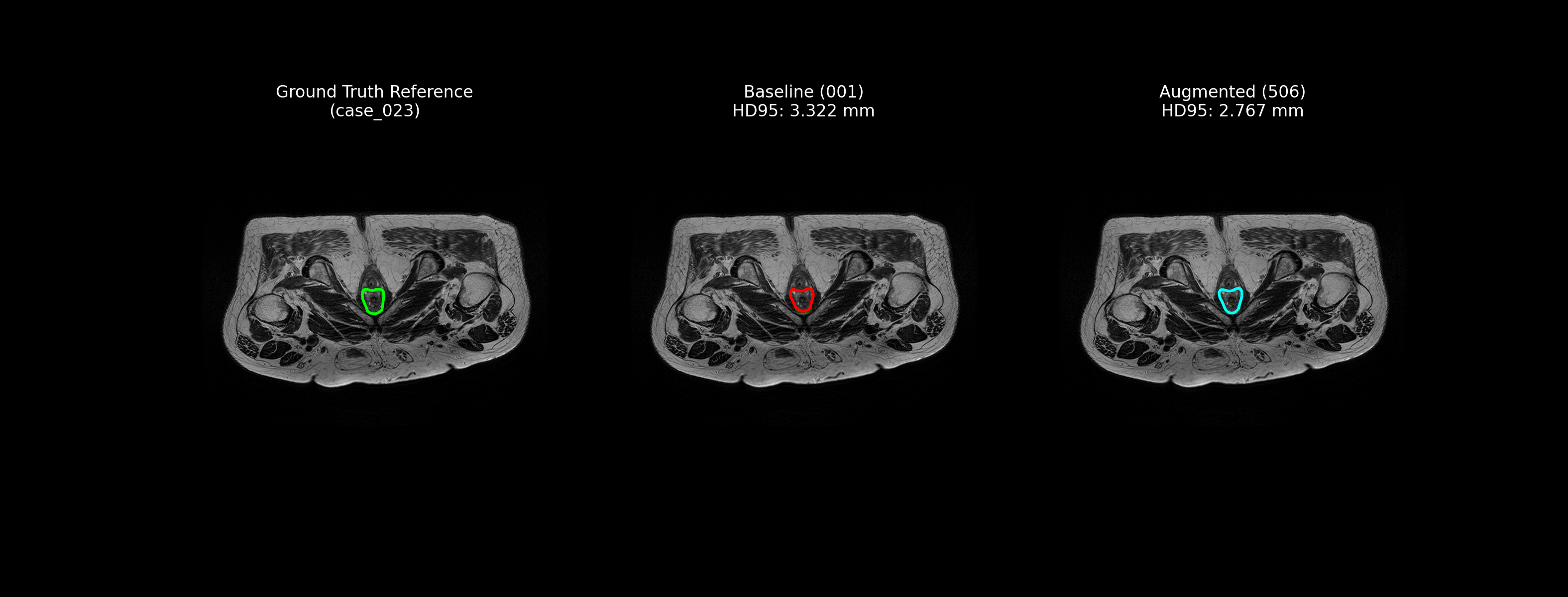}
         \caption{Case 023}
         \label{fig:hd95_case023}
     \end{subfigure}

     \begin{subfigure}[b]{0.6\textwidth}
         \centering
         \includegraphics[width=\textwidth]{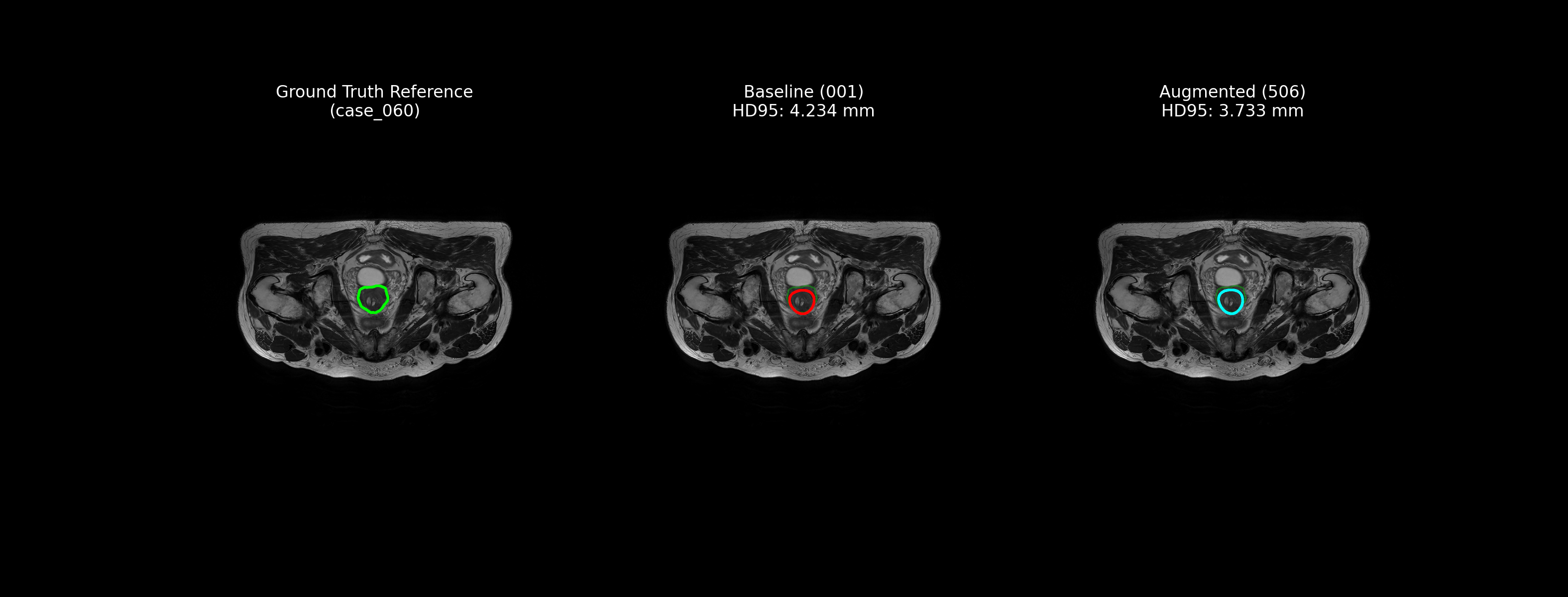}
         \caption{Case 060}
         \label{fig:hd95_case060}
     \end{subfigure}

     \begin{subfigure}[b]{0.6\textwidth}
         \centering
         \includegraphics[width=\textwidth]{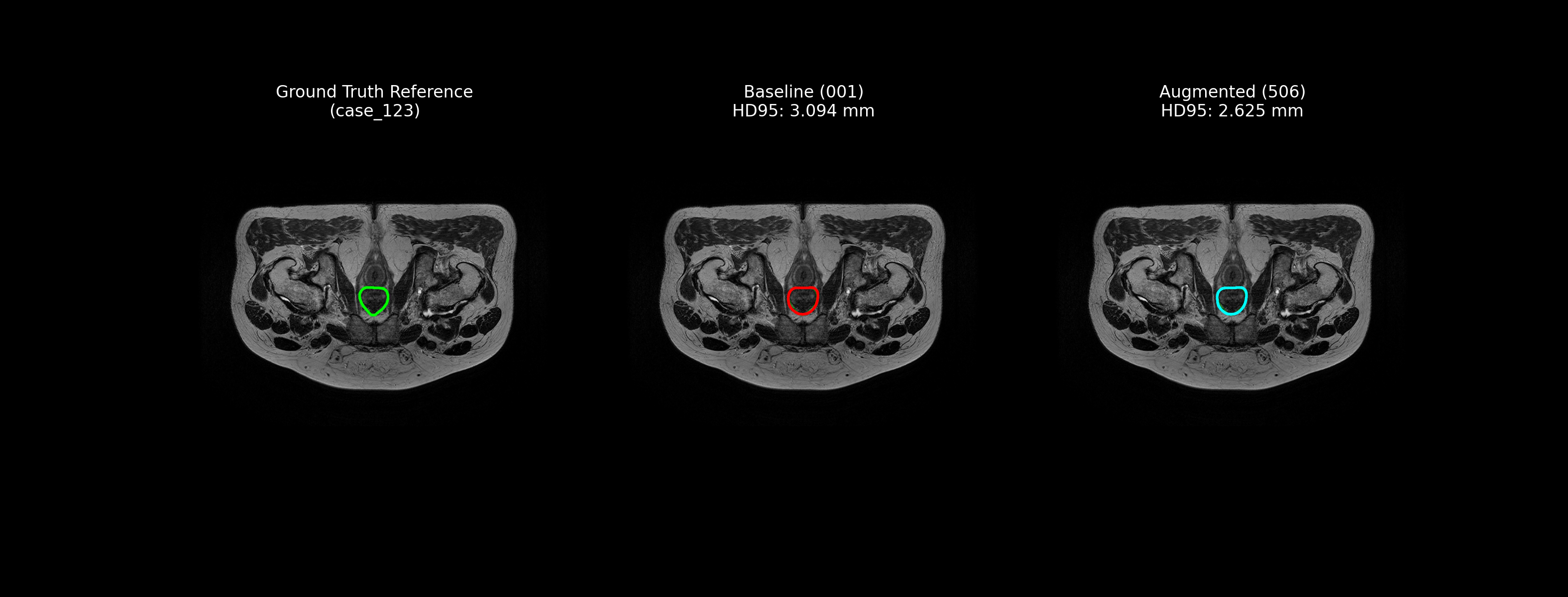}
         \caption{Case 123}
         \label{fig:hd95_case123}
     \end{subfigure}

     \caption{Qualitative comparison of boundary refinement using the 95th Percentile Hausdorff Distance (HD95). Axial T2-weighted MRI slices compare the ground truth reference (Lime, left) with the Baseline (Red, center) and Scratch (Augmented) (Cyan, right) predictions. The Augmented training strategy successfully acts as a spatial regularizer: it corrects prominent posterior leakage (a), smooths irregular lateral boundaries (b), and tightens global contours (c, d, e). These corrections translated to a reduction in maximum surface error by up to 0.764 mm in the most challenging outlier cases.}
     \label{fig:hd95_improvements_grid}
\end{figure}

The augmented (Scratch) training strategy yielded the most significant boundary corrections in cases 094, 128, 023, 060, and 123, reducing maximum surface errors by up to 0.76 mm (Figure \ref{fig:hd95_improvements_grid}). 

\end{document}